# Spin-wave gap collapse in Rh-doped $Sr_2IrO_4$

J. Bertinshaw,[1,*] J. K. Kim,[2,3,4] J. Porras,[1] K. Ueda,[1] N. H. Sung,[1] A. Efimenko,[5] A. Bombardi,[6] Jungho Kim,[4] B. Keimer,[1] and B. J. Kim[1,2,3,†]

[1]*Max Planck Institute for Solid State Research, Heisenbergstraße 1, D-70569 Stuttgart, Germany*
[2]*Department of Physics, Pohang University of Science and Technology, Pohang 790-784, South Korea*
[3]*Center for Artificial Low Dimensional Electronic Systems, Institute for Basic Science (IBS), 77 Cheongam-Ro, Pohang 790-784, South Korea*
[4]*Advanced Photon Source, Argonne National Laboratory, Argonne, Illinois 60439, USA*
[5]*ESRF–The European Synchrotron, 71 Avenue des Martyrs, F-38000 Grenoble, France*
[6]*Diamond Light Source Limited, Harwell Science and Innovation Campus, Didcot, Oxfordshire OX11 0DE, United Kingdom*

We use resonant inelastic x-ray scattering (RIXS) at the Ir $L_3$ edge to study the effect of hole doping upon the $J_{eff} = \frac{1}{2}$ Mott-insulating state in $Sr_2IrO_4$, via Rh replacement of the Ir site. The spin-wave gap, associated with XY-type spin-exchange anisotropy, collapses with increasing Rh content, prior to the suppression of the Mott-insulating state and in contrast to electron doping via La substitution of the Sr site. At the same time, despite heavy damping, the *d-d* excitation spectra retain their overall amplitude and dispersion character. A careful study of the spin-wave spectrum reveals that deviations from the $J_1$-$J_2$-$J_3$ Heisenberg used to model the pristine system disappear at intermediate doping levels. These findings are interpreted in terms of a modulation of Ir-Ir correlations due to the influence of Rh impurities upon nearby Ir wave functions, even as the single-band $J_{eff} = \frac{1}{2}$ model remains valid up to full carrier delocalization. They underline the importance of the transition metal site symmetry when doping pseudospin systems such as $Sr_2IrO_4$.

## I. INTRODUCTION

In 5*d*-electron transition metal oxides, comparable energy scales of relativistic spin-orbit coupling (SOC) and strongly correlated electron phenomena lead to enhanced quantum fluctuations and the emergence of exotic ground states [1,2]. Much of the research in this area has focused upon the spin-orbit assisted Mott-insulating ground state expressed by the layered system $Sr_2IrO_4$ [3]. Here, predominantly isotropic interactions between spin-orbit entangled magnetic moments, or pseudospins, lead to a ground state that can be largely described in terms of a Heisenberg antiferromagnet (AF) with $J_{eff} = L + S = \frac{1}{2}$ moments on a square lattice [4,5].

The close parallel between this $J_{eff}$ state and the $S = \frac{1}{2}$ high-$T_C$ parent $La_2CuO_4$ is a promising sign that $Sr_2IrO_4$ may also host unconventional superconductivity [6,7]. To what extent then does this analogy between cuprate and iridate phenomenology remain as the system is doped? While purely $J_{eff} = \frac{1}{2}$ pseudospin interactions are almost isotropic in a square lattice system, mixing with the Jahn-Teller active $J_{eff} = \frac{3}{2}$ state leads to enhanced anisotropic interactions [5]. The magnitude of splitting between the $J_{eff}$ levels in $Sr_2IrO_4$ is largely defined by the strength of SOC ($\frac{3}{2}\lambda \sim 0.6$ eV), which is smaller than the equivalent Jahn-Teller splitting of the $e_g$ states in the cuprates ($\sim 2$ eV). Indeed, it is not yet clear if the $Sr_2IrO_4$ single-band model remains valid away from the Mott state, as SOC potentially becomes ineffective in the presence of itinerant electrons [8], leading to a three-band state that breaks with cuprate behavior.

Promising signs of phenomena associated with high-$T_C$ superconductivity—Fermi arcs and a *d*-wave-like symmetry—have been identified in angle-resolved photoemission spectroscopy (ARPES) [9,10] and scanning tunneling spectroscopy [11] studies of *in situ* electron-doped $Sr_2IrO_4$. However, macroscopic evidence of superconductivity remains elusive. To date, electron doping in bulk $Sr_{2−y}La_yIrO_4$ is constrained to a maximum of $y = 0.1$ due to sample growth limitations. Up to this doping limit resonant inelastic x-ray scattering (RIXS) studies show that mostly isotropic magnetic correlations survive well into the metallic regime as long-range magnetic order is lost, properties that are similar to hole-doped cuprate physics [12–14], and hence indicative of electron-hole conjugation between the systems [7].

Nominally, hole doping through substitution of Ir with $4d^4$ Rh in $Sr_2Ir_{1−x}Rh_xO_4$ is not constrained by growth limitations. With increasing Rh content the Néel temperature ($T_N = 240$ K) systematically decreases up to the full delocalization of carriers ($x \sim 0.17$) [15]. A recent RIXS study reports that the magnon effectively hardens as holes are introduced [16], behavior that has also been noted in

---

*j.bertinshaw@fkf.mpg.de
†bjkim6@postech.ac.kr

Published by the American Physical Society under the terms of the Creative Commons Attribution 4.0 International license. Further distribution of this work must maintain attribution to the author(s) and the published article's title, journal citation, and DOI. Open access publication funded by the Max Planck Society.

electron-doped cuprates [17]. However, the doping mechanism is nontrivial and remains in question. Recent ARPES measurements [18] suggest that hybridization with Rh dopants dilutes the SOC present in the system, driving the system into a metallic state. On the other side, Ir and Rh $L_{2,3}$-edge x-ray absorption indicates that Rh impurities assume a 3+ valence [19,20], suggesting that holes are introduced in a charge compensation process. In this picture, only a relatively small fraction of Ir ions ever assumes a 5+ oxidation state, potentially as a result of disorder on the Ir site [15,20].

Here, we investigate the effect of Rh doping upon the $J_{\text{eff}} = \frac{1}{2}$ state by studying the magnetic and charge dynamics using Ir $L_3$-edge RIXS. We identify that the out-of-plane spin-wave gap, associated with anisotropic exchange interactions, collapses prior to full delocalization of the doped carriers. At the same time the dispersive character of spin-orbital excitations between the $J_{\text{eff}} = \frac{1}{2}$ and $\frac{3}{2}$ states remains unaffected. Together, these findings indicate that Rh substitution induces local changes to proximate Ir wave functions that modulate anisotropic exchange, while the Ir $J_{\text{eff}} = \frac{1}{2}$ state remains defined as the system becomes metallic.

## II. EXPERIMENTAL DETAILS

A series of single-crystal $Sr_2Ir_{1-x}Rh_xO_4$ samples with varying Rh content ($x = 0, 0.04, 0.08, 0.1,$ and $0.16$) was grown using a flux method described previously [21]. The samples were initially characterized using magnetometry and resonant x-ray diffraction [22]. RIXS measurements were conducted using beamlines ID20 at ESRF, France [23] and 27-ID at the Advanced Photon Source (APS), Argonne, IL, USA [24], in a Rowland geometry configuration with a Si(844) diced spherical analyzer, giving an overall energy resolution of $\Delta E \approx 25$ meV. Samples were mounted in a (100) × (001) scattering geometry (in orthorhombic notation) and measured at $T = 10$ K.

## III. RESULTS AND DISCUSSION

Excitation spectra for $x = 0$ and $0.1$ are shown in Figs. 1(a) and 1(b), plotted along high-symmetry directions in the magnetic Brillouin zone (BZ). The pristine sample displays a highly dispersive magnon excitation emanating from the AF zone center $(\pi, \pi)$, which extends up to around 0.2 eV at the zone boundary $(\pi, 0)$. At higher energies an additional dispersive feature lies atop the electron-hole continuum, which is attributed to $J_{\text{eff}} = \frac{1}{2} \rightarrow \frac{3}{2}$ spin-orbit excitations [25]. At $x = 0.1$ both the magnon and the spin-orbit exciton retain their dispersive character through the BZ, even as the spectrum becomes damped in the presence of charge carriers. This is notable as it shows that excitations between the Ir $\frac{1}{2}$ and $\frac{3}{2}$ manifolds remain well defined away from the strong-coupling limit.

The doping dependence of the inelastic excitations across a series of $q$ points is plotted in Figs. 1(c)–1(g). Damping increases significantly between $x = 0.04$ and $0.1$, which we note coincides with the reported crossing of the $J_{\text{eff}} = \frac{1}{2}$ band with the Fermi level [18,26]. At (0,0) in Fig. 1(c), the position of the exciton in the pristine system is marked with a dashed line. With increasing doping the exciton shifts to a smaller energy loss, indicating subtle modifications to the tetragonal crystal field $\Delta$ or $\lambda$ that define the magnitude of the $J_{\text{eff}} = \frac{1}{2}, \frac{3}{2}$ splitting [5,25]. In Figs. 1(d)–1(g) the dashed line identifies the $x = 0$ magnon. Here, while doping also leads to a significant loss of definition, only minor energy shifts from the dispersion of the pristine system are apparent. As such, within the doping regime studied, i.e., up to almost full carrier delocalization, the replacement of Ir with Rh does not significantly break the primary nearest-neighbor exchange interactions. To investigate the effect of Rh substitution upon the magnon in more detail, we conducted further measurements on a sample at an intermediate doping content where the dispersion is still defined ($x = 0.08$). A representative spectrum at $(\pi, 0)$ is plotted in Fig. 2(a), which shows a well-resolved spin wave at ∼170 meV. The spectrum is fitted with a series of Gaussian functions to account for (from left to right) the elastic line, spin wave, magnetic continuum [12], $d$-$d$ excitations, and the particle-hole continuum. The fitted magnon dispersion along high-symmetry directions is plotted in Fig. 2(b), which is compared with the undoped system measured under the same conditions, revealing small but distinct changes to the spin-wave bandwidth and dispersive behavior.

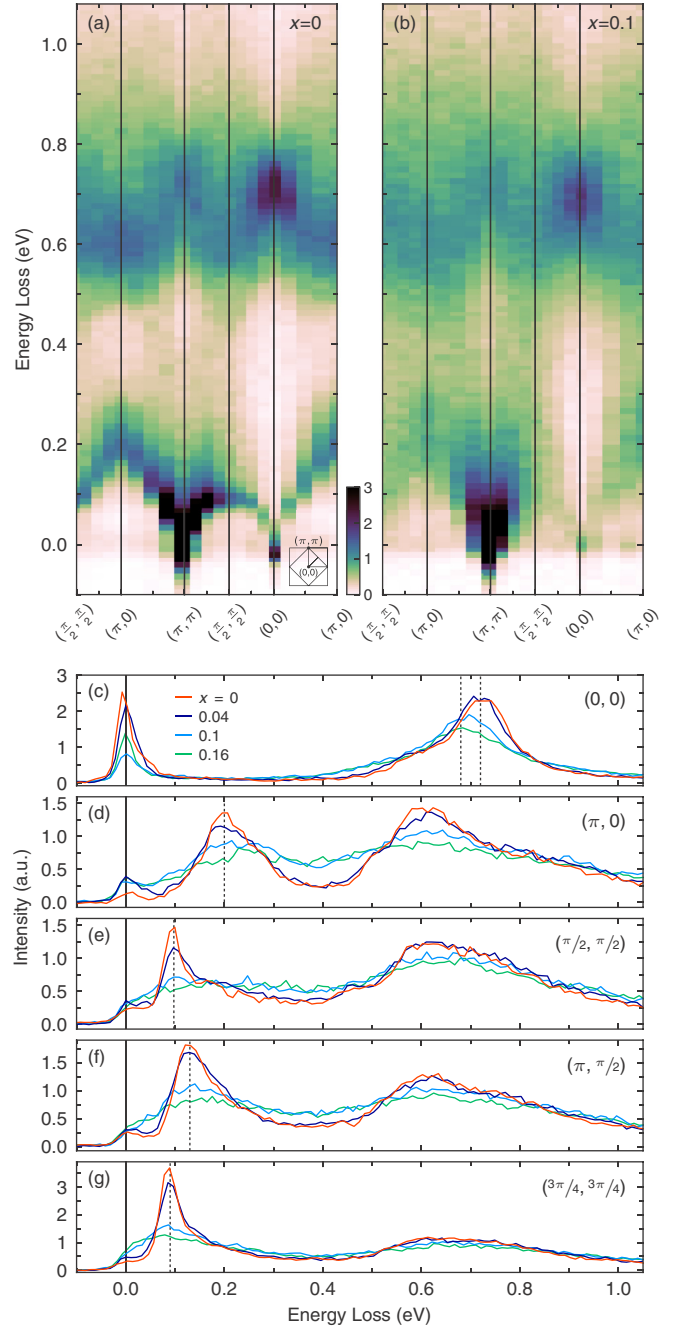

FIG. 1. (a), (b) RIXS intensity maps of $Sr_2Ir_xRh_{1-x}O_4$ ($x = 0, 0.1$) taken along high-symmetry directions at $T = 10$ K; the intensity is in arbitrary units. Inset: Magnetic BZ. (c)–(g) Comparison of RIXS spectra for $x = 0, 0.04, 0.1, 0.16$ at a series of $q$ points. The dashed line marks the $x = 0$ exciton in (c) and magnon in (d)–(g).

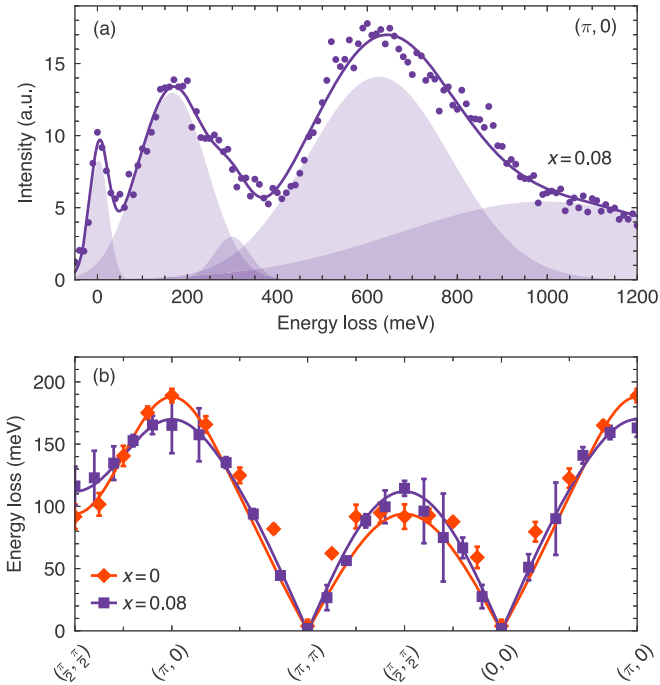

FIG. 2. (a) RIXS spectrum at $(\pi, 0)$ for $Sr_2Ir_{0.92}Rh_{0.08}O_4$ measured at $T = 10$ K. The spectral features, described in the text, are modeled with a series of Lorentzian functions. (b) The spin-wave dispersion along high-symmetry directions, extracted from the fitting method shown in (a), is compared with the pristine system. The solid lines represent a linear spin-wave model based upon the $J_1$-$J_2$-$J_3$ Heisenberg Hamiltonian.

Square lattice cuprates [27] and iridates [6] can be modeled simply in terms of a phenomenological $J_1$-$J_2$-$J_3$ Hamiltonian, which accounts for exchange interactions between respective first, second, and third nearest neighbors in the basal planes. The higher-order terms are necessary to model the downward dispersion from $(\pi, 0)$ to $(\frac{\pi}{2}, \frac{\pi}{2})$, for example. Note that anisotropic terms are not considered at this point. The parameters were tuned to reproduce the changes that arise with Rh substitution, with the resulting model dispersions plotted in Fig. 2(b) as solid lines. The introduction of Rh results in no significant change to $J_1$ [60(5) → 55(5) meV] and $J_2$ [−17(2) → −15(2) meV], while $J_3$ shows a large reduction [15(2) → 7(2) meV]. This is consistent with the fact that longer exchange paths are more sensitive to the introduction of disorder that arises from Rh substitution of the Ir site.

It is interesting to note that the $x = 0.08$ magnon is well modeled by the $J_1$-$J_2$-$J_3$ Hamiltonian, while the pristine sample dispersion shows discrepancies that are most evident around $(\pi, \pi)$ and the plateau around $(\frac{\pi}{2}, \frac{\pi}{2})$. This is seen both in Fig. 2(b) and in previous reports (e.g., Ref. [6]). The effect of Rh doping upon the extended $J_3$ term suggests that this divergence arises from longer-range exchange paths not accounted for in the $J_1$-$J_2$-$J_3$ model, likely due to the relatively wide spatial distribution of the Ir $5d$ wave function. Indeed, the spin-wave dispersion of the analogous $3d$ system $La_2CuO_4$ displays far more subtle deviations from the nearest-neighbor Heisenberg model [27].

We now turn to the impact of Rh substitution upon the exchange anisotropy present in the system. It has been previously identified that $Sr_2IrO_4$ displays a nontrivial level of easy-plane (XY-type) anisotropy [14,28]. In terms of $Sr_2IrO_4$ Heisenberg dynamics, the XY anisotropy lifts the degeneracy of in-plane and out-of-plane fluctuations. As a result, the out-of-plane spin wave becomes gapped with a magnitude reported to be on the order of 25 meV [14,28]. In order to study the effect of Rh doping upon this gap, measurements were conducted at ID20 in normal-incident geometry such that the incident $\pi$-polarized light was perpendicular to [001], maximizing scattering from out-of-plane magnetic excitations. At the same time a magnetic field of $B = 0.5$ T was applied parallel to the sample surface and in the scattering plane, using permanent magnets. This field polarized the magnetic domains along the in-plane $a$ axis, thereby suppressing sensitivity to transverse in-plane modes [29]. The aperture of the Si(844) analyzer was reduced to 2 cm in order to improve the momentum resolution at the expense of photon flux. RIXS spectra collected for the $x = 0$, 0.04, 0.1, and 0.16 samples at $(\pi, \pi) = (3, 0, 30.5)$ are shown in Fig. 3(a). Both the in-plane and out-of-plane gaps are clearly resolvable for the $x = 0$ and 0.04 samples. The in-plane mode remains due to the nonzero projection of the $a$ axis to the polarization axis in this orientation. Further measurements were collected slightly away from the magnetic zone center [22], with (3.04,0,30.5) plotted in Fig. 3(b) to emphasize the out-of-plane dispersion. For both $q$ points the $x = 0.1$ and 0.16 gap shows a significant weakening as the spectral weight shifts to lower energies, where it becomes impossible to clearly separate the in-plane and out-of plane contributions.

Two asymmetric Lorentzian functions [30] were used to model the modes in the $x = 0$ and 0.04 samples. The resulting fits are shown in Fig. 3 with the out-of-plane mode highlighted. The spectral features are resolution limited ($\Delta E = 23$ meV), and the intrinsic $Sr_2IrO_4$ out-of-plane gap lies at 34(1) meV, which is on the upper side of the previously reported range of values [14]. A small asymmetry is found [$\gamma = 0.008(1)$], which captures the magnetic continuum arising from short-range correlations [12]. The $x = 0.04$ sample shows a slight gap softening [31(1) meV], along with an increase in asymmetry [0.021(1)]. The collapse of the spin-wave gap in the $x = 0.1$ and 0.16 samples is modeled with a single asymmetric Lorentzian, which shows an increase in peak width [28.5(6) and 25.3(9) meV, respectively]. The asymmetry increases even further over the pristine system [0.036(1) for both samples], capturing the extended tail that is most clear in Fig. 3(b). We note that this asymmetry may underlie the report of an increased spin-wave gap by Clancy et al. [16], where the in-plane and out-of-plane spin-wave modes were not resolved.

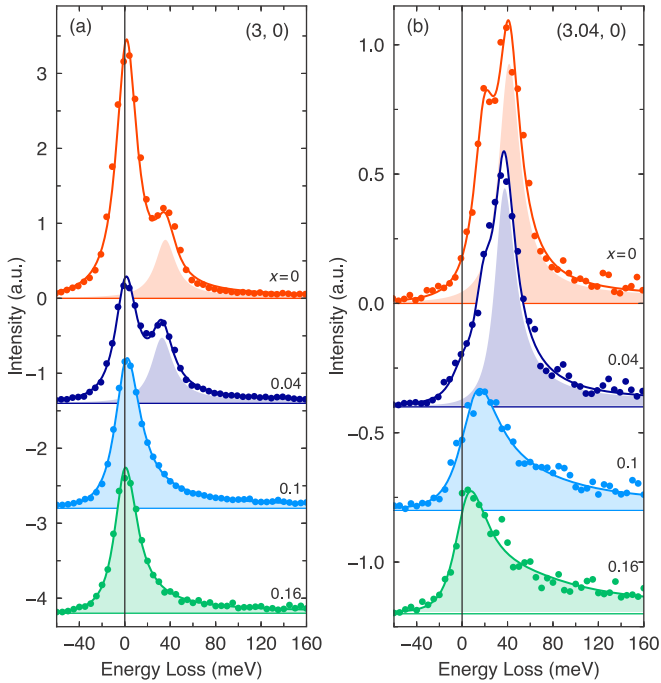

FIG. 3. (a) RIXS spectra at $T = 10$ K for $Sr_2Ir_xRh_{1-x}O_4$ ($x = 0$, 0.04, 0.1, 0.16) taken at $(\pi, \pi)$ in normal incidence and magnetic field $B = 0.5$ T to maximize the sensitivity to out-of-plane excitations. (b) Spectra taken slightly away from $(\pi, \pi)$ at (3.04,0,30.5) to resolve the dispersion of the out-of-plane spin-wave gap. $x = 0$ and $x = 0.04$ are fitted to an asymmetric bi-Lorentzian function. The peak corresponding to the out-of-plane spin wave is highlighted. $x = 0.1$ and $x = 0.16$ are modeled with a single asymmetric Lorentzian.

The excitation spectra of La- and Rh-doped $Sr_2IrO_4$ share many of the same characteristics [22]—the $J_{eff} = \frac{1}{2}$ state is preserved, while longer-range interactions are more strongly affected than nearest-neighbor coupling [12]—suggesting a common trend for carrier delocalization in $Sr_2IrO_4$ in promising parallel to cuprate physics. At the same time, carriers also have the potential to weaken the effective SOC [8], thereby softening the out-of-plane spin-wave gap as a result of reduced anisotropy in the system. However, while softening is clear with increasing Rh content, it is notably unaffected by La substitution [14]. A major difference between both doping approaches is that La is introduced via substitution of the Sr site, leaving the $IrO_2$ planes intact. This suggests that the collapse of the out-of-plane gap arises from a more direct impact of Rh impurity sites upon Ir interactions. Hybridization between Rh and Ir wave functions will lead to a more significant dilution of the effective SOC ($\lambda_{Rh} \sim 0.15$ eV). Such a condition may contribute to the decrease in the spin-orbit exciton energy from $\frac{3}{2}\lambda + \Delta \approx 0.72$ to 0.68 eV at $x = 0.16$ [Fig. 1(c)]. We note that this shift is comparable to the reported reduction in SOC determined from the ARPES dipole matrix element in a recent study by Zwartsenberg et al. [18].

Symmetry conditions will also play an important role, due to the significant orbital component of the spin-orbit entangled pseudospins. The local symmetry of each Rh site will be affected by the smaller ionic radii and differences between the 4d and 5d wave functions in terms of the spatial extension and pd covalency. A lowering of the local crystal field symmetry of Ir sites in proximity to Rh dopants is then unavoidable. The magnon out-of-plane exchange anisotropy is introduced in the spin Hamiltonian through Ising $xx$ and $yy$ bond-directional compass-type terms along the respective in-plane $a$- and $b$-oriented Ir-Ir bonds [5,28,29]. Impurity-induced changes to the crystal field will introduce different symmetry-allowed Ir-Ir exchange paths, thereby impacting the anisotropic exchange compass terms. The tetragonal crystal field will also be influenced by local changes to symmetry. Without tetragonal distortion, Ir pseudospins prefer an in-plane orientation due to the lack of compensating $zz$-type $c$ bonds within the layered perovskite crystal structure. Local changes to $\Delta$ then have the potential to affect the Ir moment orientation, strongly modifying anisotropic terms in the system [5].

The modulation of anisotropic exchange due to a distribution of compass terms, $\Delta$ and $\lambda$ scales, or a combination of these factors, is consistent with a gap softening that appears as a broad shoulder rather than as a well-defined peak in Fig. 3. Indeed, since each Rh site replaces four bonds and further affects the surrounding eight Ir-Ir bonds, it is then perhaps not surprising that the doping levels studied here have a strong effect upon the exchange anisotropy. These results therefore suggest that the Ir wave functions are affected by the Rh impurities, going beyond the effects of doping via charge compensation, with important implications for substitution of the transition metal site in pseudospin systems such as $Sr_2IrO_4$.

## IV. CONCLUSION

To summarize, our systematic RIXS measurements of the $Sr_2Ir_{1-x}Rh_xO_4$ series, from $x = 0$ up to the point of almost complete carrier delocalization at $x = 0.16$, uncovered behavior distinct from other doping approaches such as La. In this doping regime, the substitution of the 4d-electron Rh element on the Ir site has a strong but local impact upon the wave function, affecting the anisotropic exchange, which is seen as a collapse of the out-of-plane gap. We conclude by noting that as the Ir $\frac{1}{2}$ and $\frac{3}{2}$ states remain well defined in the presence of carriers, alternate doping strategies (e.g., K-ion substitution on the Sr site) may be a productive approach to uncover exotic ground states related to the intrinsic hole-doping evolution of the Ir $J_{eff} = \frac{1}{2}$ state.

## ACKNOWLEDGMENTS

We would like to thank A. Damascelli, G. Khaliullin, and H. Liu for fruitful discussions. We are grateful to S. Park, J. Kim, and H. Kim for their assistance during the RIXS measurements. We acknowledge financial support by the European Research Council under Advanced Grant No. 669550 (Com4Com). This work was supported by IBS-R014-A2. J.B. was supported by the Alexander von Humboldt Foundation. J.K.K. was supported by the Global Ph.D. Fellowship Program by National Research Foundation of Korea (Grant No. 2018H1A2A1059958). The use of the Advanced Photon Source at Argonne National Laboratory was supported by the U.S. DOE under Contract No. DE-AC02-06CH11357.